\makeatletter\let\ifGm@compatii\relax\makeatother
\documentclass[pre,aps,twocolumn,floatfix,superscriptaddress]{revtex4}
\usepackage{eso-pic,calc}
\usepackage{graphicx}
\usepackage{amsmath, amsthm, amssymb}
\usepackage{epsfig}
\usepackage{latexsym}
\usepackage{bm}

\def\beqr{\begin{eqnarray}}
\def\eqnr{\end{eqnarray}}
\def\beq{\begin{equation}}
\def\bc{\begin{center}}
\def\ec{\end{center}}
\def\eqn{\end{equation}}

\def\prl#1#2#3{{ Phys. Rev. Lett.} {\bf #1}, #2 (#3)}

\def\pre#1#2#3{Phys. Rev. E {\bf #1}, #2 (#3)}

\def\pnas#1#2#3{Proc. Natl. Acad. Sci. (USA) {\bf #1}, #2 (#3)}

\def\physa#1#2#3{Physica A {\bf #1}, #2 (#3)}

\def\nat#1#2#3{Nature {\bf #1}, #2 (#3)}

\def\sc#1#2#3{Science {\bf #1}, #2 (#3)}

\begin{document}
\title{Correspondence between noisy sample space reducing process and records in correlated random events}
\author{Avinash Chand Yadav}
\affiliation{Department of Physics \& Astronomical Sciences, Central University of Jammu, Samba 181 143, India}

\begin{abstract}
We study survival time statistics in a noisy sample space reducing (SSR) process. Our simulations suggest that both the mean and standard deviation scale as $\sim N/N^{\lambda}$, where $N$ is the system size and $\lambda$ is a tunable parameter that characterizes the process. The survival time distribution has the form $\mathcal{P}_{N}(\tau)\sim N^{-\theta}J(\tau/N^{\theta})$, where $J$ is a universal scaling function and $\theta = 1-\lambda$. Analytical insight is provided by a conjecture for the equivalence between the survival time statistics in the noisy SSR process and the record statistics in a correlated time series modeled as drifted random walk with Cauchy distributed jumps. 
\end{abstract}

\maketitle
\section{Introduction}
The explanation of the origin of power--law or scale--invariant features associated with complex systems has been an active and fascinating area of research in statistical physics. This feature is well recognized in natural science as Zipf's law \cite{Zipf_1949} that states how frequently a word occurs in a given typical text is inversely proportional to its rank such as $P(i) \sim i^{-\lambda}$, with $\lambda$ close to 1.  Typical instances range from the size-distributions of cities \cite{Makse_1995}, commercial firms \cite{Axtell_2001}, and fluctuations in financial markets \cite{Gabaix_2003}, to diameters of moon craters \cite{Neukum_1994} (see Ref.~\cite{Newman_2005} for more examples). However, in many situations the exponent is not exactly 1, but it takes value in [0, 1]. Although expressing frequency versus rank plot is not a preferred way in statistical physics, the frequency distribution also has a decaying power law dependence, with the two exponents being related; the exponent for frequency distribution is $\gamma = (1+1/\lambda)$ \cite{Newman_2005}.

Earlier proposed mechanisms  that explain scale--invariant features \cite{Newman_2005} include self-organized criticality \cite{Bak_1987, Bak_1996, Dhar_2006}, preferential attachment \cite{Albert_1999}, a combination of exponentials \cite{Newman_2005}, an inverse of quantities \cite{Sornette_2002}, and multiplicative processes \cite{Montroll_1982, Sornette_1998,  Amir_2012}.
Recently, state or sample space reducing (SSR) stochastic process \cite{Murtra_2015} has been proposed as an independent route to explain such features, particularly, Zipf's law.  The SSR mainly reflects the process of aging in complex systems, where the size of state space, namely the set of all possible allowed states, reduces as time advances.  Striking examples of SSR include the process of sentence formation in linguistics \cite{Murtra_2015}, fragmentation of materials \cite{Krapivsky_1994}, polymerization process \cite{Krapivsky_2011}, and diffusion on weighted, directed and acyclic graphs \cite{Thurner_2016}. The sample space may not be strictly reducing in some cases, but one can incorporate the occasional expansion as a fluctuation; this is termed as noisy SSR process.

A highly fruitful strategy that seems relevant here is that a mapping between seemingly different problems in statistical physics not only facilitates computation but can also provide deeper insight \cite{Moloney_2016}. Some interesting examples studied with this point of view are the following: A mapping of percolation on trees and Brownian excursions \cite{Moloney_2016}, or the directed abelian sandpile on a narrow strip and a random walk on a ring \cite{Yadav_2012} or the fragmentation and aggregation process in polymerization and the study of cycles in random permutation with uniform measure \cite{Krapivsky_2011}. In the present context, a mapping between survival time statistics for SSR processes and records statistics of independent and identically distributed (iid) random variables \cite{Yadav_2016} is most pertinent.

In this work, our focus is on the {\em noisy} SSR process. We first review the recent developments on the SSR process; in Ref.~\cite{Yadav_2016} the statistics of survival time (the life span) of the SSR process was studied. Exact results, supported by simulations, showed that both mean and variance of the survival time vary in a logarithmic manner as a function of system size, and the asymptotic probability distribution is Gaussian. The correspondence noted here, between the survival time statistics for the SSR process and the records statistics of iid random variables, not only provides a deeper insight about the system, but it also becomes apparent to identify other applications of the SSR process, for example, polymerization process \cite{Krapivsky_2011} to cycles statistics in random permutation  \cite{Flajolet_2009}.

We show here that the records statistics in a random time series, suitably modeled as a correlated random process (such as random walk) is potentially equivalent to noisy SSR process, and can be completely characterized by the exponent $\lambda$.  Studies of records statistics are relevant in many applications such as financial time series or price fluctuations of a commodity that are examples of a correlated noisy signal.

The organization of the paper is as follows. The definition of the noisy SSR process with detailed numerical results for the survival time statistics is presented in Sec.~\ref{sec_2}. Subsequently, analytical and numerical results for the records statistics in correlated random events are studied in Sec.~\ref{sec_3}. Finally, our results are summarized in Sec.~\ref{sec_4} which also includes a discussion.

\section{\label{sec_2}The survival time statistics for the noisy SSR process}
We begin with recalling the definition of SSR stochastic process \cite{Murtra_2015}. This can be visualized as a directed random hopping process on $\mathbb{Z}^+$, the set of positive integers \cite{Yadav_2016}. Let us denote the instantaneous state of the process by $x(t) = K$, where $1\leq K \leq N$; or in terms of a random walk, the walker is at $K$th site after time $t$. The dynamical update rules are the following: At the next time step $t+1$, the walker can go to any site in the interval $[1, K-1]$ with equal probability, i.e., $1/(K-1)$. As a result of this jump, if the outcome is $L$ such that $1< L \leq K-1$, then the walker can similarly jump to any site in the interval $[1, L-1]$ with probability $1/(L-1)$. The same process is repeated until the walker reaches to site 1. The boundary conditions are $x(0) = N$ and $x(\tau) = 1$, and the life span or survival time of the process is a discrete random variable such that $1\leq\tau \leq N$.  Clearly, the process is strictly reducing. Since $x = 1$ represents that there is only one state for the system under consideration, this would act as an absorbing state. One available state means that the system's state does not change as a function of time, reflecting no dynamics.

Further, if the restriction on the direction of motion for the SSR process is relaxed, this results in an unconstrained random hopping. Here, the dynamics is such that walker can jump from the present site to any other site with uniform probability. Then, the noisy SSR process can be described as a superposition of SSR and unconstrained random hopping, where the SSR process is executed with probability $\lambda$.  The state of the noisy SSR process is evolved in discrete time as 
\beq
x_{t+1} = G_{\lambda}(x_{t}) = \begin{cases} G_1(x_t),~{\rm with~probability~}\lambda,\\  G_0(x_t),~{\rm with~probability~}1-\lambda,\end{cases} \nonumber
\eqn 
with $\lambda\in[0,1 ]$ and the functions $G_1, G_0$, and $G_{\lambda}$ denote the update rules for strictly SSR, unconstrained hopping process, and noisy SSR, respectively. Clearly, for $\lambda = 1$ the standard SSR process is recovered, while $\lambda = 0$ corresponds to unconstrained random hopping. In this case, the survival time is not necessarily be restricted by $N$ as observed for the SSR process.

Since for the SSR process, the total number of all possible allowed configurations for fixed $N$ is finite $2^{N-1}$, directly using combinatorial methods the exact analytical results can be provided in a straightforward manner for several observables such as survival time, visiting, and occupation probability distribution \cite{Yadav_2016}. For a single realization of the SSR process, a site may be visited once. Thus the visiting probability distribution $P(i)$, i.e., the probability with which $i$th site is visited, and the occupation probability $p(i)$ that denotes on an average how many times the site is visited are related by a normalization factor as $P(i) = p(i)/p(1)$ \cite{Murtra_2015}. 

\begin{figure}[t]
  \centering
  \scalebox{0.65}{\includegraphics{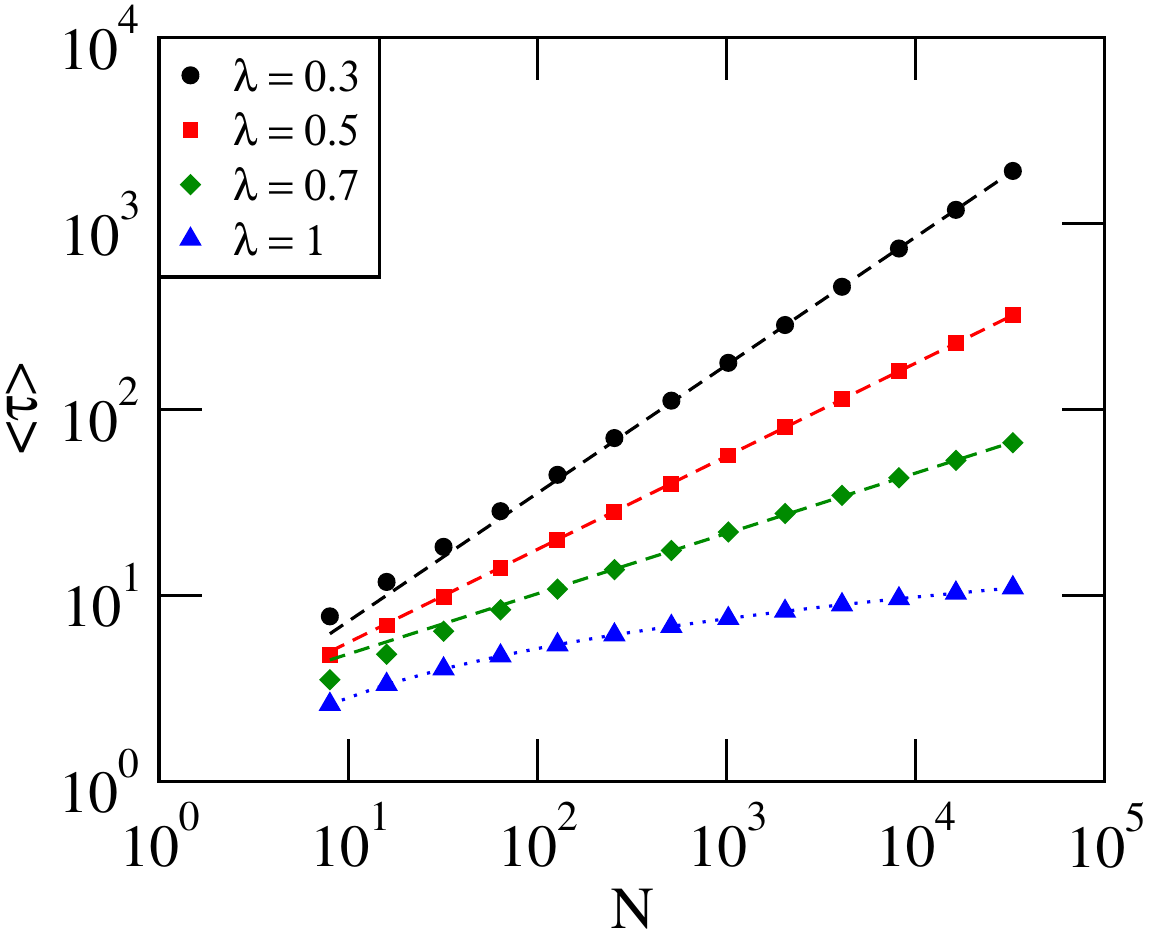}}
   \caption{The mean survival time as a function of the system size for different values of the parameter $\lambda$. Broken lines are best-fit curves, and for $\lambda = 1$ the behaviour is exactly logarithmically varying shown with dotted line. Here, the maximum run time of an individual realization is $10^4$, and the total number of independent realizations of the process is $10^6$. $N$ is taken in steps $2^j$ with $j = 3$ to 15. Also, when $\lambda = 0.3, 0.5,$ and  0.7, the estimated exponents are $\theta = 0.69\pm 0.01, 0.501\pm 0.001,$ and  $0.32\pm 0.01$, respectively. }
\label{fig1}
\end{figure}

\begin{figure}[t]
  \centering
  \scalebox{0.65}{\includegraphics{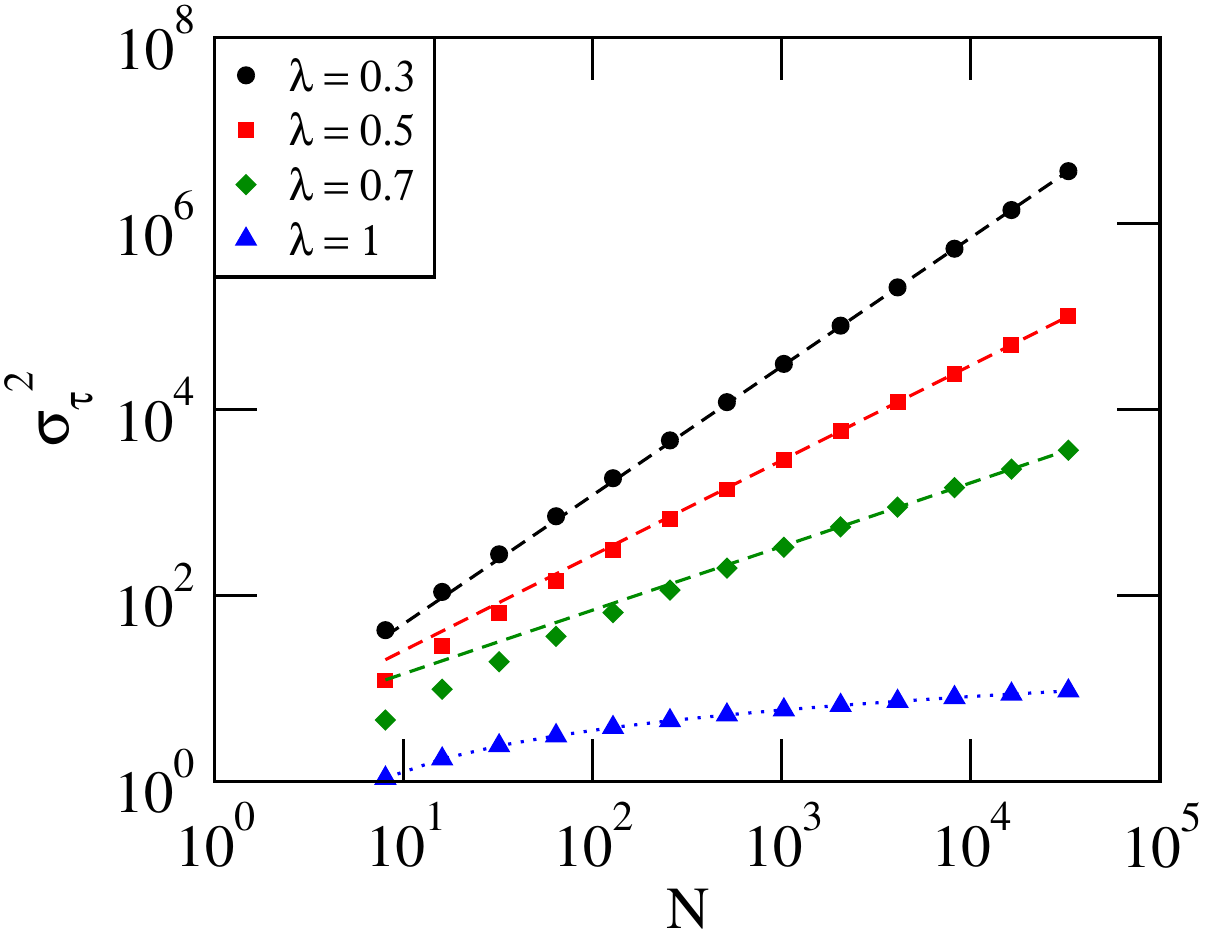}}
   \caption{The variance of survival time as a function of the system size for different values of the parameter $\lambda$. For different values of $\lambda$ = 0.3, 0.5, and 0.7, the corresponding estimated exponents are $1.39\pm 0.01, 1.02\pm 0.01,$ and $0.68\pm 0.02$, respectively.}
\label{fig2}
\end{figure}

On the other hand, the case of noisy SSR process is such that here the aforesaid approach does not work for computing the probability distribution of an observable since the total number of all possible configurations is not finite. This is a consequence of the fact that a site may be visited several times for a single realization of the process. Although the noisy SSR process can survive for a long time, the process is stationary  as it is bounded, i.e., $1\le x \le N$. Also, this is a Markov process that is a future state only depends on the present state, not on the past states. These properties can be utilized to obtain an expression for the occupation probability \cite{Murtra_2015}, given by 
\beq
p(i) \sim i^{-\lambda}.
\eqn
 This is basically Zipf's law.

\begin{figure}[t]
  \centering
  \scalebox{0.65}{\includegraphics{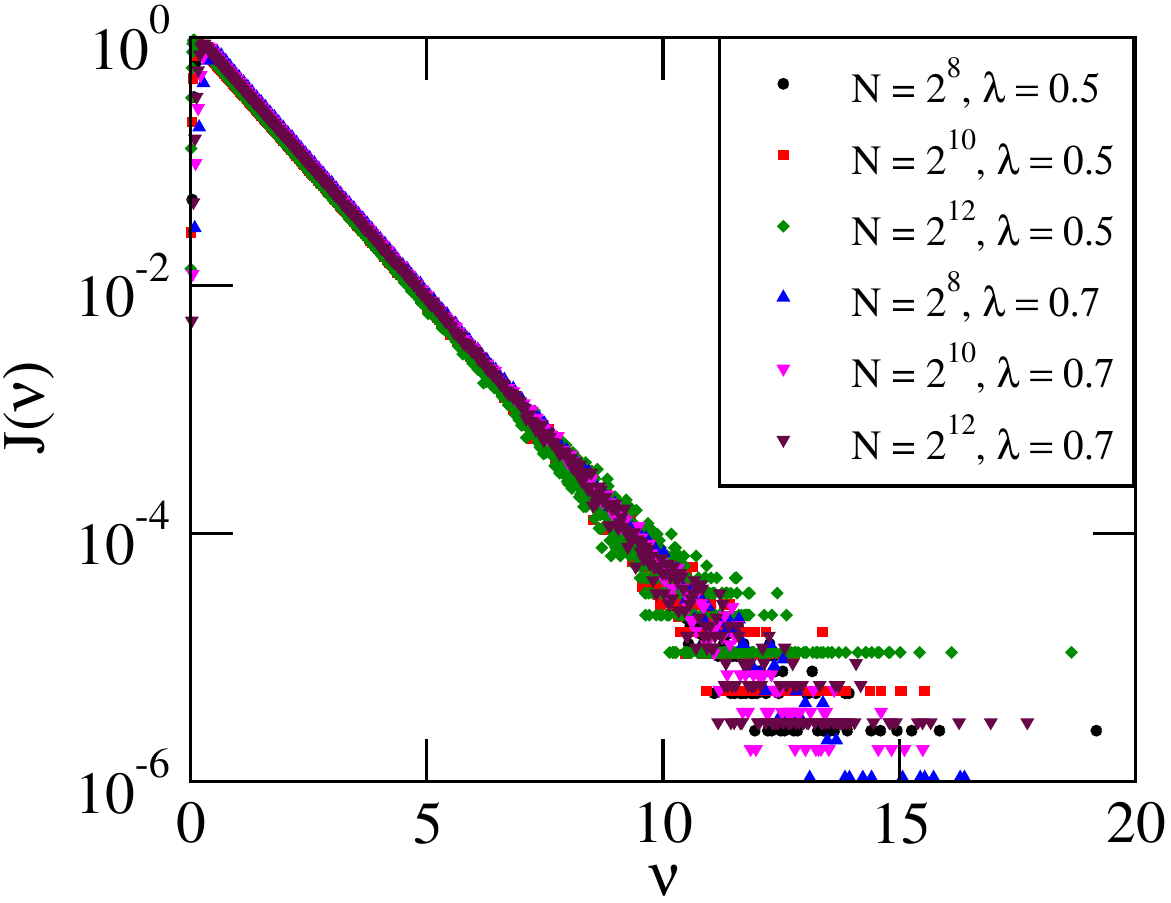}}
   \caption{The universal scaling function for the survival time distribution. Here, the total number of independent realizations of the process is $10^7$. }
\label{fig3}
\end{figure}

\begin{figure}[t]
  \centering
  \scalebox{0.65}{\includegraphics{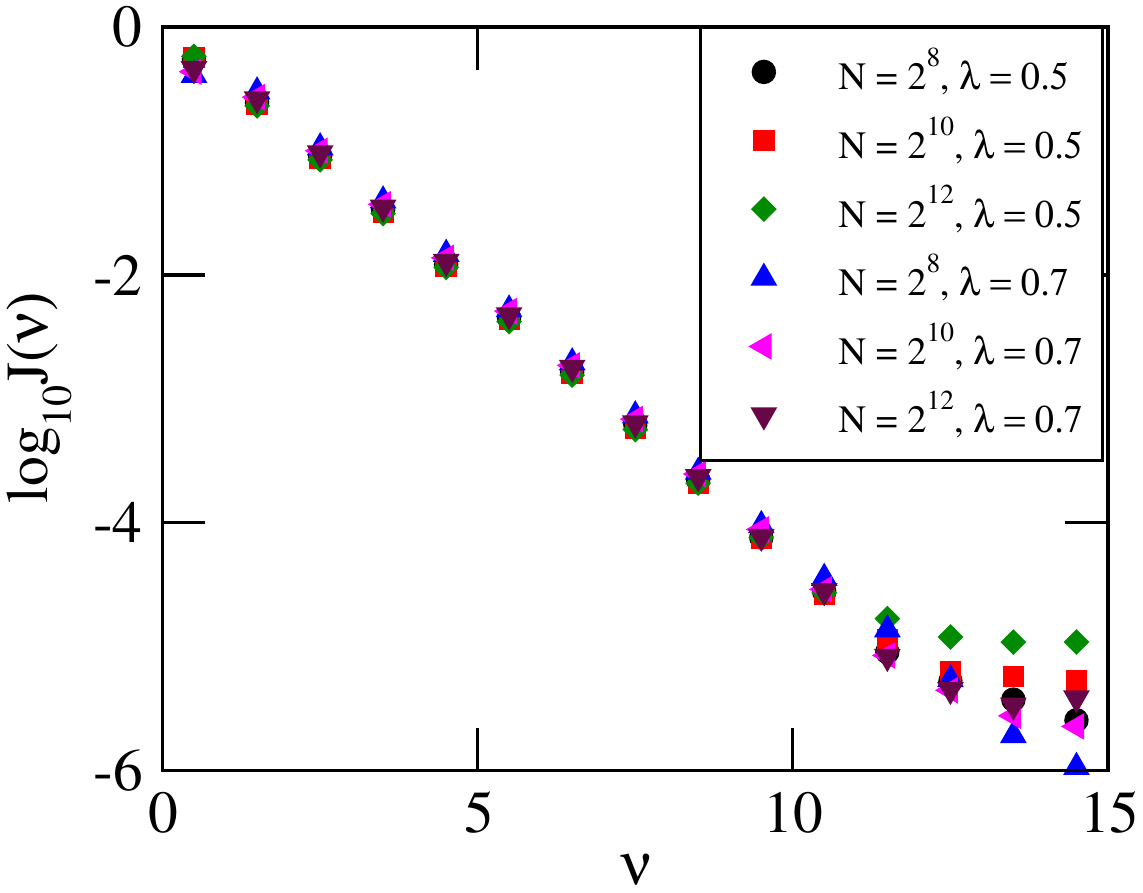}}
   \caption{The same curve as in Fig.~\ref{fig3}, but log-linear binned data, with unit bin width. $\nu$ is chosen as midpoint of the bin.}
\label{fig4}
\end{figure}

We first present numerical results obtained by performing simulations using Monte Carlo method.  Results, shown in Figs.~\ref{fig1} and \ref{fig2} suggest that the mean and variance of the survival time as a function of the system size,  within the statistical error, follow simple scaling forms
\beq
\langle \tau \rangle \sim N^{\theta},
\label{av_tau_1}
\eqn
and 
\beq
\sigma_{\tau}^2 \sim N^{2\theta},
\eqn
with $\theta = 1-\lambda$. This relation will be derived below. 
Further, the survival time distribution for the noisy SSR process exhibits a universal scaling behavior independent of $N$ and $\lambda$, when plotted with scaled argument variable $\nu = \tau/N^{\theta}$ of a function $J(\nu) = N^{\theta}\mathcal{P}_{N}(\tau)$ [see Figs.~\ref{fig3} and \ref{fig4}]. This implies that the survival time distribution can be expressed as 
\beq
\mathcal{P}_{N}(\tau) \sim N^{-\theta}J\left(\frac{\tau}{N^{\theta}}\right),
\eqn
where the universal scaling function has an exponentially decaying form such as 
\beq
J(\nu) \sim \exp(-\beta\nu),
\eqn
where $\beta$ is a constant. Clearly, $J(\nu)\to0$ for large $\nu$. 

In order to ensure that there is an exact correspondence between two processes [as discussed below], the complete statistics should behave similarly. For this reason, it becomes important to numerically check not only both the mean and variance, but the entire distribution of the survival time for the noisy SSR process.

To gain an analytical insight for the average survival time, let us introduce an indicator variable $\eta_k(t)$. When the noisy SSR process visits $k$th site after time $t$, $\eta_k(t) = 1$, and 0 otherwise. As the same site may be visited several times, we write $\zeta_k = \sum_{t} \eta_k(t)$, where $\zeta_k$ denotes how many times $k$th site is visited for a single realization.  By the definition of occupation probability, we have $\langle \zeta_k\rangle = p(k)$. As the total number of visits over the entire space, starting from $N$ and eventually reaching at site 1, is the survival time,  the mean survival time can be computed as $\langle \tau \rangle = \sum_{k=1}^{N}\langle \zeta_k \rangle$.  Clearly, the mean survival time is proportional to the area under occupation distribution. For large $N$, treating the sum $\sum_k$ as integration,  we have  
\beq
\langle \tau\rangle = \int_{1}^{N}p(k)dk \sim  \int_{1}^{N}\frac{dk}{k^{\lambda}} \sim N^{1-\lambda}.
\label{eq_tau}
\eqn
Comparing Eqs.~(\ref{av_tau_1}) \& (\ref{eq_tau}), we find a relation between the exponents $\theta$ and $\lambda$,
\beq
\theta = 1 - \lambda.
\label{lambda}
\eqn

\section{\label{sec_3}The records statistics in a correlated time series}
Consider a discrete time symmetric random walk with a constant drift term \cite{Wergen_2012, Wergen_2013} described as
\beq
X(t+1) = X(t) + c + \xi(t),
\label{eq1}
\eqn
where $c$ denotes the drift and  $\xi$ is a random variable. {When $\xi$ follows L\'evy stable distribution or distributed with a probability density function, namely $f_{\alpha}(\xi) \sim |\xi|^{-1-\alpha}$ for large $\xi$ with $0<\alpha<2$, Eq.~(\ref{eq1}) describes drifted {\em L\'evy flight}, a class of non-Gaussian Markov random processes. Since the variance of $\xi$ diverges, consequently long jumps may occur. Here, we are particularly interested in the case  $\alpha = 1$ that corresponds to} Cauchy distribution, with explicit density function given by
\beq
f(\xi) = \frac{1}{\pi}\frac{1}{1+\xi^2}.
\label{eq2}
\eqn
Set the initial condition $X(0) = 0$, and the process is executed upto time $t = N$. In order to generate, random variables with density function $f(\xi)$, first generate uniformly distributed random variables in the range $u \in [-\pi/2, \pi/2]$, and then apply a nonlinear transformation, namely $\xi = \tan{u}$. Note that the dynamical equation for $X(t)$ describes standard Brownian motion (SBM) when $c = 0$ and $\xi$ being Gaussian distributed random variables. 

\begin{figure}[b]
  \centering
  \scalebox{0.6}{\includegraphics{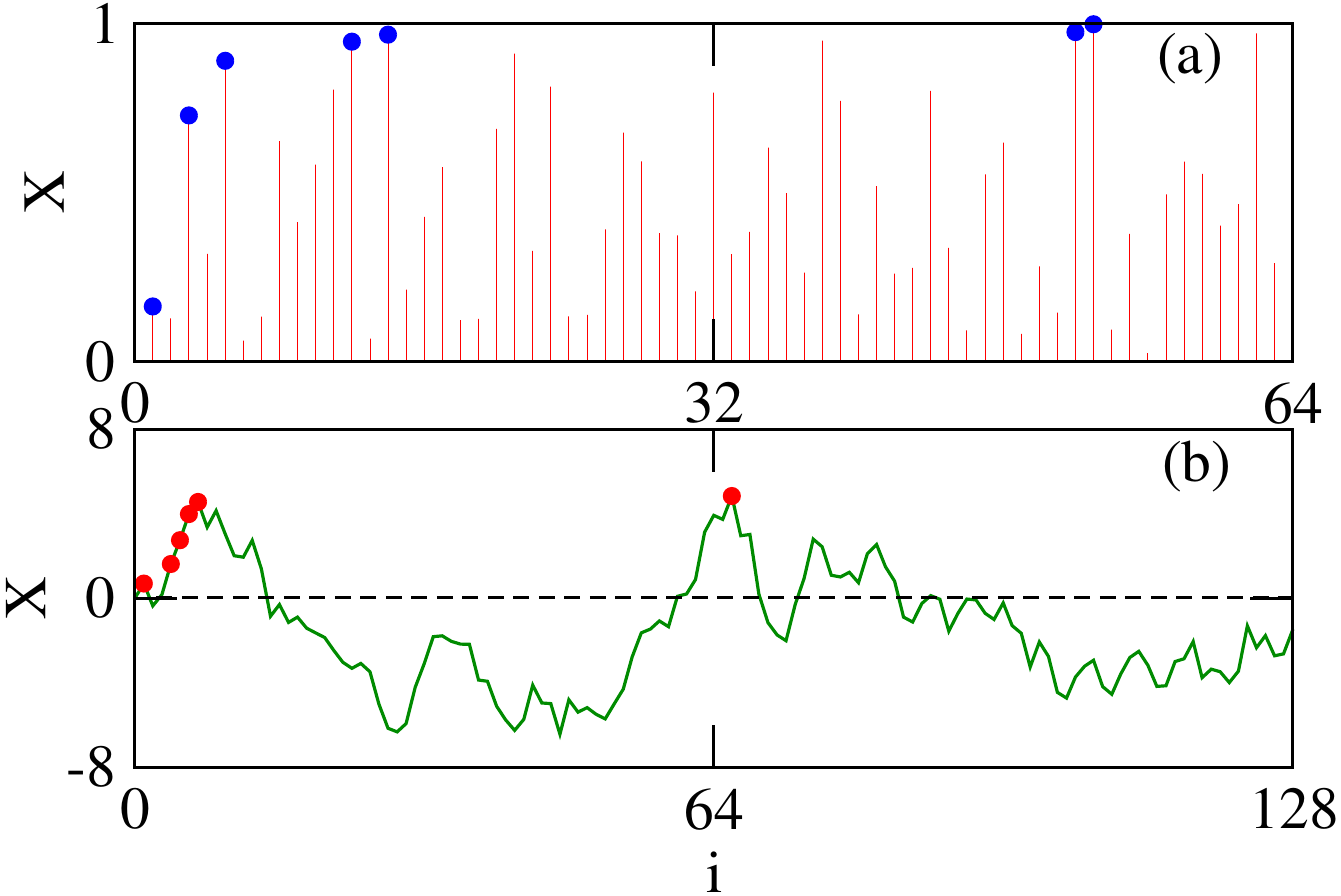}}
   \caption{Typical examples of records formation in noisy time series modeled as (a) iid random variables with uniform distribution in the interval [0, 1] and (b) a correlated time series describing SBM. Shaded circles denote record events.}
\label{fig5}
\end{figure}

Let us recall the definition of record for a sequence of random events, with the total number of events $N$. An event forms a record if it is the largest or smallest with respect to all previously existing entries [see Fig.~\ref{fig5}]. As it has been established that the total number of records follows a Gaussian distribution, with the mean number of records and variance both varying logarithmically with argument $N$, if $X(t)$'s are iid random variables.  If the event is modeled by Eqs.~(\ref{eq1}) \& (\ref{eq2}), it has been found that the mean number of records grows algebraically as $\langle R\rangle \sim N^{\theta(c)}$ with
\beq
\theta(c) = \frac{1}{2} + \frac{1}{\pi}\arctan(c),
\eqn
where $c \in [-\infty, \infty]$ that leads to $\theta \in [0, 1]$ \cite{Doussal_2009}. In addition, the probability distribution satisfies a scaling behavior $\mathcal{P}_{N}(R) \sim N^{-\theta}J(R/N^{\theta})$, with the variance $\sigma_{R}^2 \sim N^{2\theta}$.

\begin{figure}[t]
  \centering
  \scalebox{0.65}{\includegraphics{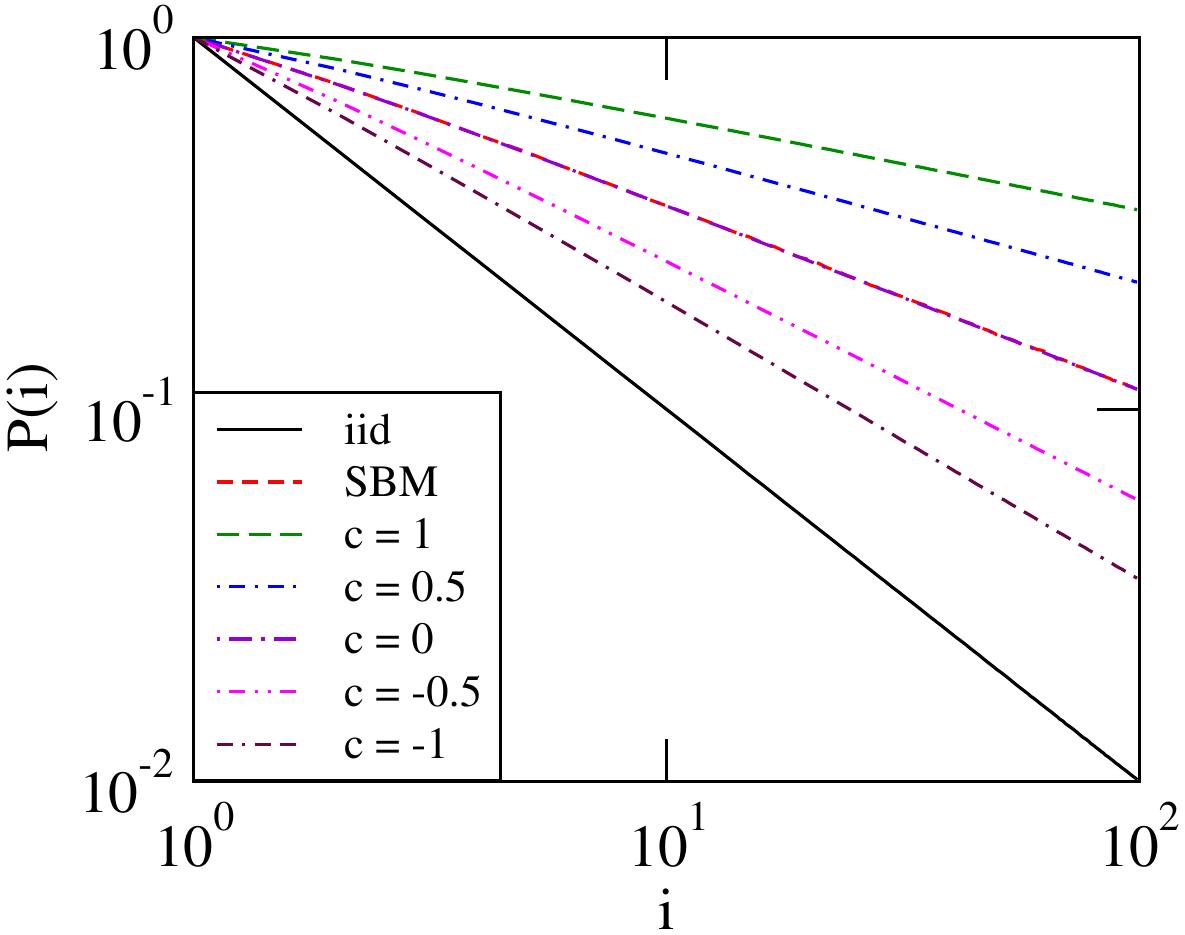}}
   \caption{Zipf's law: The plot of the probability that $i$th event forms  a record for different stochastic processes. Here, $N = 10^2$, and $10^8$ independent realizations of the process.}
\label{fig6}
\end{figure}

Based on above observations, we conjecture that there is an equivalence between the survival time statistics in the noisy SSR process and the records statistics in random events modeled by Eq.~(\ref{eq1}) \& (\ref{eq2}), although a rigorous proof for this equivalence seems a challenging problem. However, if the conjecture is valid, we can further say that the probability  that  $i$th event forms a record, where $1\le i\le N$, is equivalent to how frequently that corresponding site is visited by the noisy SSR process, namely, the visiting probability distribution. Here, this is expressed as
\beq
P(i) \sim i^{-\lambda(c)}.
\label{vf_rc}
\eqn
Since for this process each event may form a record once in a single realization, the visiting and occupation probability distributions are directly related. The analysis [Eqs.~(\ref{eq_tau}) \& (\ref{lambda})] done in the Sec.~II shows that $\lambda(c) = 1 - \theta(c)$. Numerical results shown in Fig.~\ref{fig6} agree well with this observation.

When $c$ is positive and large,  consequently each event is likely to surpass the previous event; the probability that $k$th event forms a record becomes uniform or $(\lambda\to 0)$. On the other extreme, when $c$ is negative and small, a record is less likely to be broken, while the corresponding probability tends to vary as $1/k$ since each event would behave independently and occur with equal probability.

Unlike the indicator variable earlier used for the noisy SSR process in Sec.~\ref{sec_2}, similarly, we here introduce such a variable $\bar{\eta}_k$,
\begin{equation}
\bar{\eta}_k = \begin{cases}1,~~~~~{\rm if}~X(k)~{\rm is~a~record,}\\  0,~~~~~{\rm otherwise.} \end{cases}
\end{equation} 
With this indicator variable, the total number of records can be easily computed as $R = \sum_{k=1}^{N}\bar{\eta}_k$. Further, the mean records is simply $\langle R \rangle = \sum_{k=1}^{N}\langle\bar{\eta}_k \rangle$, where $\langle \bar{\eta}_k \rangle = P(k)$. Then, using Eq.~(\ref{vf_rc}), it can be verified that $\langle R\rangle \sim N^{\theta(c)}$, with $\theta(c) = 1 - \lambda(c)$. This provides a subtle evidence that statistically records for events described Eqs.~(\ref{eq1}) \& (\ref{eq2}) are equivalent to survival time in the noisy SSR process.

Since $X(t=0) = 0$ is the initial condition, only positive events would subsequently form record. For iid random variables [see Fig.~\ref{fig5}(a)], the first event forms a record with unit probability. For the correlated time series modeled by Eq.~(\ref{eq1}) without drift, the first event can be either positive or negative with equal probability. Therefore, the first event forms a record with 1/2 probability. 

Although there are several stochastic processes studied under the field {\em records statistics}, the case that we considered is the most useful, since it is simple and able to explain the full spectrum of the exponent for the Zipf's law, as observed for the noisy SSR process. In a recent Ref.~\cite{Aliakbari_2017}, the records for fractional Brownian motion and fractional Gaussian noise have been studied. We note that the records statistics in these cases also behave in a similar manner to that of  the survival time statistics for the noisy SSR or strictly SSR process, respectively. The Hurst exponent, a characteristic of  the fractional
Brownian motion, plays the role of the exponent $1-\lambda$. Such a stochastic processes have been used to model correlated events found to occur in a diverse physical systems, suggesting potentially wide applicability for the noisy SSR. It should be noted that  the different exponent  values for the Zipf's law correspond to different {\em universality classes}.

\section{\label{sec_4}Conclusion}

We have studied the survival time statistics for a noisy SSR process. Based on simulation results, we find that the mean and standard deviation both satisfy a simple scaling behavior with the system size as $\sim N^{\theta}$, with $\theta = 1- \lambda$. In addition, the survival time probability distribution has the form $\mathcal{P}_{N}(\tau)\sim N^{-\theta}J(\tau/N^{\theta})$, where the universal scaling function shows an exponentially decaying feature.  These results further enable us to conjecture for the existence of a map between the survival time statistics in noisy SSR process and the records statistics in a class of correlated events such as  a symmetric drifted random walk with Cauchy distributed jumps. Established analytical results for the records statistics thereby provide an enhanced understanding of the noisy SSR process. Also, the record formation in correlated random events explains the Zipf's law, where the exponent can be continuously tuned by directly changing the value of drift. 

As noted that the SSR process has correspondence with records statistics in iid random variables, and the records statistics, in this case, can be mapped with statistics of cycles in random permutation with uniform measure \cite{Yadav_2016}. Also, polymerization that consists of fragmentation and aggregation processes can be mapped to the formation of cycles in random permutation with uniform measure \cite{Krapivsky_2011}. It is suggestive, using transitivity, that the SSR process can apply to such situations as well. 

\section*{ACKNOWLEDGMENTS}

ACY thanks R. Ramaswamy for critical reading of the manuscript and providing useful comments.


\begin{thebibliography}{99}
%zipf law
\bibitem{Zipf_1949} G. K.  Zipf, {\it Human Behavior and the Principle of Least Effort} (Addison-Wesley,
Reading, MA, 1949).

% distribution of city
\bibitem{Makse_1995} H. A. Makse, S. Havlin, and H. E. Stanley, \nat{377}{608}{1995}.

% Firm size 
\bibitem{Axtell_2001} R. L. Axtell, Science {\bf 293}, 1818 (2001).

% fluctuations in financial market
\bibitem{Gabaix_2003} X. Gabaix, P. Gopikrishnan, V. Plerou, and H. E. Stanley, \nat{423}{510}{2003}.

% crater diameter dist.
\bibitem{Neukum_1994} G. Neukum and B. A. Ivanov, {\it Hazards Due to Comets and Asteroids} (University of Arizona Press, Tucson, 1994).

%power law
\bibitem{Newman_2005} M. E. J. Newman, Contemporary Physics {\bf 46}, 323 (2005).


% SOC
\bibitem{Bak_1987} P. Bak, C. Tang, and K. Wiesenfeld, \prl{59}{381}{1987}.
\bibitem{Bak_1996} P. Bak, {\it How Nature Works: The Science of Self Organized Criticality} (Copernicus Press, New York, 1996).
\bibitem{Dhar_2006} D. Dhar, \physa{369}{29}{2006}.

% Preferential attachement 
\bibitem{Albert_1999} A.-L. Barab{\'a}si and R. Albert, \sc{286}{509}{1999}.

% Inverse of quantities: Mechanism for powerlaws without selforganization.
\bibitem{Sornette_2002} D. Sornette, Int. J. Mod. Phys. C {\bf 13}, 133 (2002).

%Multiplicative process
\bibitem{Montroll_1982} E. W. Montroll and M. F. Shlesinger, Proc. Natl. Acad. Sci. U. S. A. {\bf 79}, 3380 (1982); B. J. West and M. F. Shlesinger, Int. J. Mod. Phys. B {\bf 3}, 795b (1989).
\bibitem{Sornette_1998} D. Sornette, \pre{57}{4811}{1998}.
\bibitem{Amir_2012} A. Amir, Y. Oreg, and Y. Imry, \pnas{109}{1850}{2012}.

% ssr pnas
\bibitem{Murtra_2015} B. Corominas-Murtra, R. Hanel, and S. Thurner, \pnas{112}{5348}{2015}.


% fragmentation
\bibitem{Krapivsky_1994} P. L. Krapivsky and E. Ben-Naim, \pre{50}{3502}{1994}.

%Kinetics of ring formation
\bibitem{Krapivsky_2011} E. Ben-Naim and P. L. Krapivsky, \pre{83}{061102}{2011}.

% SSR2
\bibitem{Thurner_2016} B. Corominas-Murtra, R. Hanel, and S. Thurner, New J. Phys. {\bf 18}, 093010 (2016).

% Role of map/correspondence
\bibitem{Moloney_2016} F. Font-Clos and N. R. Moloney, Phys. Rev. E {\bf 94}, 030102 (2016).

% sandpile model
\bibitem{Yadav_2012} A. C. Yadav, R. Ramaswamy, and D. Dhar, \pre{85}{061114}{2012}.

%survival time in ssr
\bibitem{Yadav_2016} A. C. Yadav, \pre{93}{042131}{2016}.

% cycles in random permutation
\bibitem{Flajolet_2009} P. Flajolet and R. Sedgewick, {\it Analytic combinatorics}, (Cambridge University
Press, Cambridge, 2009).

%Records in stochastic processes - Theory and applications
%\bibitem{Wergen_2013} G. Wergen, arXiv:1211.6005v2.
\bibitem{Wergen_2012} S. N. Majumdar, G. Schehr, G. Wergen, J. Phys. A: Math. Theor. {\bf 45}, 355002 (2012). 
%\bibitem{Schehr_2013} G. Schehr and S. N. Majumdar, arXiv:1305.0639v1.
\bibitem{Wergen_2013} G. Wergen, J. Phys. A: Math. Theor. {\bf 46}, 223001 (2013).


%Drifted  RW Cauchy distributed jump
\bibitem{Doussal_2009} P. Le Doussal and K. J. Wiese, \pre{79}{051105}{2009}.

\bibitem{Aliakbari_2017} A. Aliakbari, P. Manshour, and M. J. Salehi, arXiv:1704.04377. 

\end{thebibliography}
\end{document}